\documentclass[10pt,conference]{IEEEtran}
\IEEEoverridecommandlockouts

\usepackage{cite}
\usepackage{kotex}
\usepackage{float}
\usepackage{amsmath,amssymb,amsfonts}
\usepackage[ruled, vlined]{algorithm2e}
\usepackage{graphicx}
\usepackage{textcomp}
\usepackage{xcolor}
\usepackage{listings}
\usepackage{caption}
\usepackage{subcaption}
\usepackage{multirow}
\usepackage{booktabs}
\usepackage{makecell}

\definecolor{dkgreen}{rgb}{0,0.6,0}
\definecolor{gray}{rgb}{0.5,0.5,0.5}
\definecolor{mauve}{rgb}{0.58,0,0.82}
\lstdefinestyle{myScalastyle}{
  frame=tb,
  language=scala,
  aboveskip=3mm,
  belowskip=3mm,
  showstringspaces=false,
  columns=fixed,
  basicstyle={\footnotesize\ttfamily},
  numbers=none,
  keywordstyle=\color{blue},
  commentstyle=\color{dkgreen},
  stringstyle=\color{mauve},
  frame=single,
  breaklines=true,
  breakatwhitespace=true,
  tabsize=3,
}
\lstdefinestyle{smallScalastyle}{
  frame=tb,
  language=scala,
  aboveskip=3mm,
  belowskip=3mm,
  showstringspaces=false,
  columns=fixed,
  basicstyle={\scriptsize\ttfamily},
  numbers=none,
  keywordstyle=\color{blue},
  commentstyle=\color{dkgreen},
  stringstyle=\color{mauve},
  frame=single,
  breaklines=true,
  breakatwhitespace=true,
  tabsize=3,
}

\lstdefinestyle{ires}{
  frame=tb,
  aboveskip=3mm,
  belowskip=3mm,
  showstringspaces=false,
  columns=fixed,
  basicstyle={\footnotesize\ttfamily},
  numbers=none,
  keywordstyle=\color{blue},
  commentstyle=\color{dkgreen},
  stringstyle=\color{mauve},
  frame=single,
  breaklines=true,
  breakatwhitespace=true,
  tabsize=3,
}

\lstdefinelanguage{JavaScript}{
  keywords={async, await, break, case, catch, class, const, continue, debugger,
    default, delete, do, else, enum, export, extends, false, finally, for,
    function, if, import, in, instanceof, new, null, return, super, switch, this,
    throw, true, try, typeof, var, void, while, with, yield},
  keywordstyle=\color{blue}\bfseries,
  ndkeywordstyle=\color{darkgray}\bfseries,
  identifierstyle=\color{black},
  sensitive=false,
  comment=[l]{//},
  morecomment=[s]{/*}{*/},
  commentstyle=\color{purple}\ttfamily,
  stringstyle=\color{red}\ttfamily,
  morestring=[b]',
  morestring=[b]"
}

\lstdefinestyle{myJSstyle}{
  language=JavaScript,
  extendedchars=true,
  basicstyle=\footnotesize\ttfamily,
  showstringspaces=false,
  showspaces=false,
  numbers=none,
  tabsize=2,
  breaklines=true,
  showtabs=false,
  captionpos=b
}

\newcolumntype{?}{!{\vrule width 1pt}}

\newcommand{\code}[1]{\text{\lstinline[style=myJSstyle]!#1!}}

\newcommand{\norm}[1]{||{#1}||}

\newcommand{\mytextsf}[1]{\textsf{\small #1}}

\newcommand{\tool}{\mytextsf{JEST}}
\newcommand{\jiset}{\mytextsf{JISET}}

\newcommand{\telem}[2]{\multicolumn{1}{#1}{\text{#2}}}
\newcommand{\telembf}[2]{\multicolumn{1}{#1}{\textbf{#2}}}
\newcommand{\telemsf}[2]{\multicolumn{1}{#1}{\mytextsf{#2}}}

\newcommand{\ruleset}{\mathbb{R}}
\newcommand{\worklist}{W}

\begin{document}

\title{$\textsf{JEST}$: $N\!+\!1$-version Differential Testing of\\ Both JavaScript Engines and Specification}

\author{
  \IEEEauthorblockN{Jihyeok Park}
  \IEEEauthorblockA{\textit{School of Computing} \\
  \textit{KAIST}\\
  Daejeon, South Korea\\
  jhpark0223@kaist.ac.kr}

  \and

  \IEEEauthorblockN{Seungmin An}
  \IEEEauthorblockA{\textit{School of Computing} \\
  \textit{KAIST}\\
  Daejeon, South Korea\\
  h2oche@kaist.ac.kr}

  \and

  \IEEEauthorblockN{Dongjun Youn}
  \IEEEauthorblockA{\textit{School of Computing} \\
  \textit{KAIST}\\
  Daejeon, South Korea\\
  f52985@kaist.ac.kr}

  \and

  \IEEEauthorblockN{Gyeongwon Kim}
  \IEEEauthorblockA{\textit{School of Computing} \\
  \textit{KAIST}\\
  Daejeon, South Korea\\
  gyeongwon.kim@kaist.ac.kr}

  \and

  \IEEEauthorblockN{Sukyoung Ryu}
  \IEEEauthorblockA{\textit{School of Computing} \\
  \textit{KAIST}\\
  Daejeon, South Korea\\
  sryu.cs@kaist.ac.kr}
}

\maketitle

\begin{abstract}
  Modern programming follows the continuous integration (CI) and continuous
  deployment (CD) approach rather than the traditional waterfall model.  Even
  the development of modern programming languages uses the CI/CD approach to
  swiftly provide new language features and to adapt to new development
  environments.  Unlike in the conventional approach, in the modern CI/CD
  approach, a language specification is no more the oracle of the language semantics
  because both the specification and its implementations (interpreters or compilers) can co-evolve.
  In this setting, both the specification and implementations may have bugs, and
  guaranteeing their correctness is non-trivial.

  In this paper, we propose a novel \textit{$N$+1-version differential testing} to resolve
  the problem.  Unlike the traditional differential testing, our approach
  consists of three steps: 1) to automatically synthesize programs guided by the
  syntax and semantics from a given language specification, 2) to generate
  conformance tests by injecting assertions to the synthesized programs to check their final program
  states, 3) to detect bugs in the specification and implementations via executing
  the conformance tests on multiple implementations, and 4) to localize bugs on
  the specification using statistical information.  We actualize our approach for
  the JavaScript programming language via \( \tool \), which performs
  $N$+1-version differential testing for modern JavaScript engines and ECMAScript,
  the language specification describing the syntax and semantics
  of JavaScript in a natural language.  We evaluated \( \tool \) with four JavaScript engines that
  support all modern JavaScript language features and the latest version of
  ECMAScript (ES11, 2020).  \( \tool \) automatically synthesized 1,700
  programs that covered 97.78\% of syntax and 87.70\% of
  semantics from ES11.  Using the assertion-injected JavaScript programs,
  it detected 44 engine bugs in four different engines and 27
  specification bugs in ES11.
\end{abstract}


\section{Introduction}\label{sec:intro}

In Peter O'Hearn's keynote speech in ICSE 2020, he quoted the following from
Mark Zuckerberg's Letter to Investors~\cite{mzletter}:
\begin{quote}
  The Hacker Way is an approach to building that involves continuous improvement
  and iteration.  Hackers believe that somethings can always be better, and that
  nothing is ever complete.
\end{quote}
Indeed, modern programming follows the continuous integration (CI) and
continuous deployment (CD) approach~\cite{cicd} rather than the traditional waterfall model.
Instead of a sequential model that divides software development into
several phases, each of which takes time, CI/CD amounts to a cycle of
quick software development, deployment, and back to development with
feedback. Even the development of programming languages uses the CI/CD approach.

Consider JavaScript, one of the most widely used programming languages
for client-side and server-side programming~\cite{nodejs} and
embedded systems~\cite{moddable,espruino,tessel2}.  Various JavaScript
engines provide diverse extensions to adapt to fast-changing user demands.  At
the same time, ECMAScript, the official specification that describes the syntax and
semantics of JavaScript, is annually updated since ECMAScript 6 (ES6,
2015)~\cite{es6} to support new features in response to user demands.
Such updates in both the specification and implementations in tandem make it
difficult for them to be in sync.

Another example is Solidity~\cite{officialSolDoc}, the standard smart contract programming language
for the Ethereum blockchain.  The Solidity language specification is continuously
updated, and the Solidity compiler is also frequently released.  According to
Hwang and Ryu~\cite{solidity-gap}, the average number of days between consecutive
releases from Solidity 0.1.2 to 0.5.7 is 27.  In most cases, the Solidity compiler reflects
updates in the specification, but even the specification is revised
according to the semantics implemented in the compiler.  As in JavaScript,
bidirectional effects in the specification and the implementation make
it hard to guarantee their correspondence.

In this approach, both the specification and the implementation may contain bugs,
and guaranteeing their correctness is a challenging task.
The conventional approach to build a programming language is uni-directional from
a language specification to its implementation.  The specification is believed to
be correct and the conformance of an implementation to the specification is
checked by dynamic testing.  Unlike in the conventional approach, in the modern CI/CD
approach, the specification may not be the oracle, because both the
specification and the implementation can co-evolve.

In this paper, we propose a novel \textit{$N$+1-version differential testing}, which
enables testing of co-evolving specifications and their implementations.  The
differential testing~\cite{diff-test} is a testing technique, which executes $N$
implementations of a specification concurrently for each input, and detects a
problem when the outputs are in disagreement.  In addition to $N$
implementations, our approach tests the specification as well using a
mechanized specification.  Recently, several approaches to extract syntax and
semantics directly from language specifications are presented\cite{jiset,
extract-x86, extract-arm}.  We utilize them to bridge the gap between
specifications and their implementations through conformance tests generated
from mechanized specifications.  The $N$+1-version differential testing consists of
three steps: 1) to automatically synthesize programs guided by the syntax and
semantics from a given language specification, 2) to generate conformance tests
by injecting assertions to the synthesized programs to check their final program
states, 3) to detect bugs in the specification and implementations
via executing the conformance tests on multiple implementations, and 4) to
localize bugs on the specification using statistical information.

Given a language specification and $N$ existing real-world
implementations of the specification, we
automatically generate a conformance test suite from the specification with
assertions in each test code to make sure that the result of running the code
conforms to the specification semantics.  Then, we run the test suite for $N$
implementations of the specification.  Because generated tests strictly comply
with the specification, they reflect specification errors as well, if any.  When
one of the implementations fails in running a test, the
implementation may have a bug, as in the differential testing.  When
most of the implementations fail in running a test, it is highly likely that
the specification has a bug.  By automatically generating a rich set of test
code from the specification and running them with implementations of the
specification, we can find and localize bugs either in the specification written
in a natural language or in its implementations.

To show the practicality of the proposed approach, we present $\tool$, which
is a \underline{J}avaScript \underline{E}ngines and
\underline{S}pecification \underline{T}ester using $N$+1-version differential testing.
We implement $\tool$ by extending $\jiset$~\cite{jiset}, a JavaScript
IR-based semantics extraction toolchain, to utilize the syntax and semantics
automatically extracted from ECMAScript.  Using the extracted syntax,
our tool automatically synthesizes initial seed programs and expands the program
pool by mutating specific target programs guided by semantics coverage.  Then,
the tool generates conformance tests by injecting assertions to synthesized
programs.  Finally, $\tool$ detects and localizes bugs using execution
results of the tests on $N$ JavaScript engines.  We evaluate our tool with four
JavaScript engines (Google V8\cite{v8}, GraalJS\cite{graaljs}, QuickJS\cite{qjs},
and Moddable XS\cite{xs}) that support all modern JavaScript language
features and the latest ECMAScript (ES11, 2020).

The main contributions of this paper include the following:
\begin{itemize}
  \item Present \textit{$N$+1-version differential testing}, a novel solution to the new
    problem of co-evolving language specifications and their implementations.
  \item Implement $N$+1-version differential testing for JavaScript engines and
    ECMAScript as a tool called $\tool$. It is the first tool that automatically generates conformance
    tests for JavaScript engines from ECMAScript.  While the coverage
    of Test262, the official conformance tests, is 91.61\% for statements
    and 82.91\% for branches, the coverage of the conformance tests generated by the tool
    is 87.70\% for statements and 78.30\% for branches.
  \item Evaluate $\tool$ with four modern JavaScript engines and the latest
    ECMAScript, ES11.  Using the generated conformance test
    suite, the tool found and localized 44 engine bugs in four different
    engines and 27 specification bugs in ES11.
\end{itemize}

\section{$N$+1-version Differential Testing}\label{sec:idea}
This section introduces the core concept of $N$+1-version differential testing with a
simple running example.  The overall structure consists of two
phases: a conformance test generation phase and a bug detection and localization
phase.

\begin{figure}[t]
  \centering
  \begin{subfigure}[t]{0.48\textwidth}
    \includegraphics[width=\textwidth]{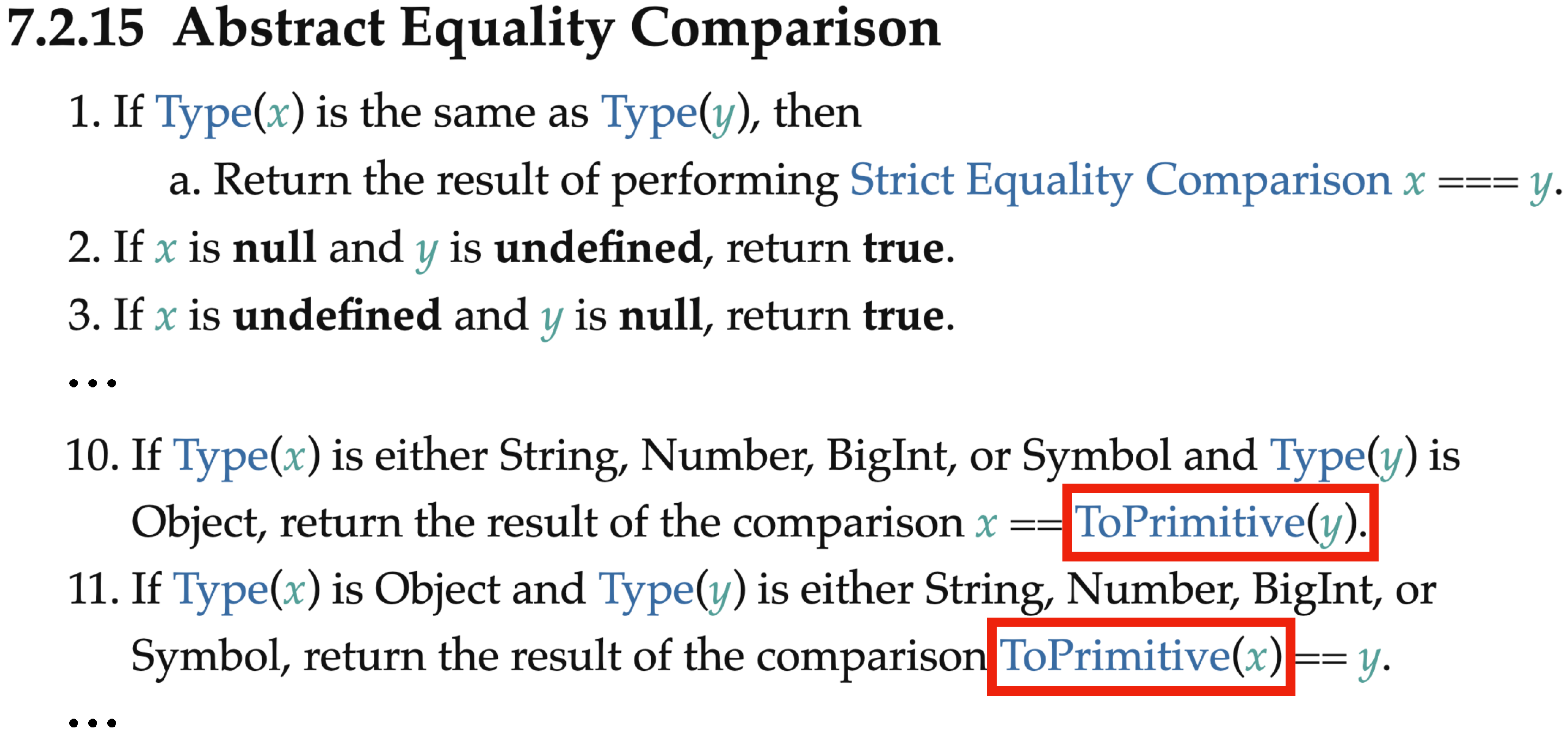}
\vspace*{-.5em}
    \caption{The \textbf{Abstract Equality Comparison} abstract algorithm in
    ES11}
    \label{fig:example-algo}
  \end{subfigure}
  \begin{subfigure}[t]{0.43\textwidth}
    \begin{lstlisting}[style=myJSstyle]
// JavaScript engines: exception with "err"
// ECMAScript (ES11) : result === false
var obj = { valueOf: () => { throw "err"; } };
var result = 42 == obj;
    \end{lstlisting}
\vspace*{-.5em}
    \caption{JavaScript code using abstract equality comparison}
    \label{fig:example-js}
  \end{subfigure}
  \begin{subfigure}[t]{0.45\textwidth}
    \begin{lstlisting}[style=myJSstyle]
try {
  var obj = { valueOf: () => { throw "err"; } };
  var result = 42 == obj;
  assert(result === false);
} catch (e) {
  assert(false);
}
    \end{lstlisting}
\vspace*{-1em}
    \caption{JavaScript code with injected assertions}
    \label{fig:example-injected}
  \end{subfigure}
  \caption{Abstract algorithm in ES11 and code example using it}
  \label{fig:example}
  \vspace*{-1em}
\end{figure}

\subsection{Main Idea}

Differential testing utilizes the cross-referencing oracle, which is
an assumption that any discrepancies between program behaviors on the same input
could be bugs.  It compares the execution results of a program with the same
input on $N$ different implementations.  When an implementation produces a different
result from the one by the majority of the implementations, differential testing
reports that the implementation may have a bug.

On the contrary, $N$+1-version differential testing utilizes not only the cross-referencing oracle
using multiple implementations but also a mechanized specification.  It first
generates test code from a mechanized specification, and tests 
$N$ different implementations of the specification using the generated test code
as in differential testing.  In addition, it can detect possible bugs in the specification
as well when most implementations fail for a test.  In such cases,
because a bug in the specification could be triggered by the test, it
localizes the bug using statistical information as we explain later in this section.

\begin{figure*}[t]
  \centering
  \includegraphics[width=0.98\textwidth]{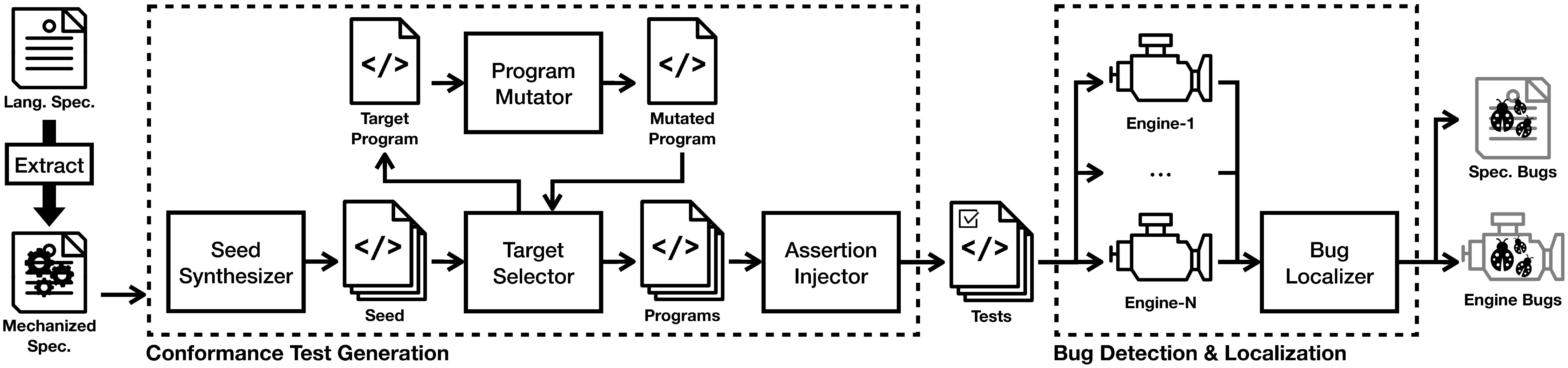}
  \vspace*{.5em}
  \caption{Overall structure of $N$+1-version differential testing for $N$
    implementations (engines) and one language specification}
  \label{fig:overall}
  \vspace*{-1em}
\end{figure*}

\subsection{Running Example}

We explain how $N$+1-version differential testing works with a simple
JavaScript example shown in Figure~\ref{fig:example}.

Figure~\ref{fig:example}(a) is an excerpt from ECMAScript 2020 (ES11),
which shows some part of the \textbf{Abstract Equality Comparison} abstract algorithm.
It describes the semantics of non-strict equality comparison such as \code{==} and \code{\!=}.
For example, $\code{null == undefined}$ is $\code{true}$ because of
the algorithm step 2.  According to the steps 10 and 11, if the type of a value is
String, Number, BigInt, or Symbol, and the type of the other value is Object, the algorithm calls
\textbf{ToPrimitive} to convert the JavaScript object to a primitive value.
Note that this is a specification bug caused by unhandled abrupt completions!
To express control diverters such as exceptions, $\code{break}$, $\code{continue}$,
$\code{return}$, and $\code{throw}$ statements in addition to normal values,
ECMAScript uses ``abrupt completions.''
ECMAScript annotates the question mark prefix (?) to all function
calls that may return abrupt completions to denote that they should be checked.
  However, even though \textbf{ToPrimitive} can
produce an abrupt completion, the calls of \textbf{ToPrimitive} in steps 10 and
11 do not use the question mark, which is a bug.

Now, let's see how $N$+1-version differential testing can detect the
bug in the specification. Consider the example JavaScript code in
Figure~\ref{fig:example}(b), which triggers the above specification bug.
In the \textbf{Abstract Equality Comparison} algorithm, variables
\textit{x} and \textit{y} respectively denote $\code{42}$ and an object with a property
named $\code{valueOf}$ whose value is a function throwing an error.
Step 10 calls \textbf{ToPrimitive} with the object as its argument, and the call returns
an abrupt completion because the call of $\code{valueOf}$ throws an error.
However, because the call of \textbf{ToPrimitive} in step 10 does not
use the question mark, the specification semantics silently ignores the abrupt completion and
returns $\code{false}$ as the result of comparison. Using the specification semantics,
we can inject assertions to check that the code does not throw any errors as shown
in Figure~\ref{fig:example}(c). Then, by running the code with the injected assertions
on $N$ JavaScript engines, which throw errors, we can find that the specification
may have a bug.  Moreover, we can localize the bug using statistical information:
because most conformance tests that go through steps 10 and 11 of the algorithm
would fail in most of JavaScript engines, we can use the information
to localize the bug in the steps 10 and 11 of \textbf{Abstract
Equality Comparison} with high probability.

\subsection{Overall Structure}

Figure~\ref{fig:overall} depicts the overall structure of $N$+1-version differential testing
for $N$ different implementations (engines) and one language specification.
It takes a mechanized specification extracted from a given language
specification, it first performs the conformance test generation phase, which
automatically generates conformance tests that reflect the language
syntax and semantics described in the specification.  Then, it performs the
bug detection and localization phase, which detects and localizes bugs
in the engines or the specification by comparing the results
of the generated tests on $N$ engines.

The functionalities of each module in the overall structure are as follows:

\subsubsection{Seed Synthesizer}
The first module of the conformance test generation phase is \mytextsf{Seed Synthesizer},
which synthesizes an initial seed programs using the language syntax.
Its main goal is to synthesize (1) a few number of
(2) small-sized programs (3) that cover possible cases in the syntax rules as many as possible.

\subsubsection{Target Selector}
Starting from the seed programs generated by \mytextsf{Seed Synthesizer}
as the initial \emph{program pool}, \mytextsf{Target Selector}
selects a target program in the program pool that potentially increases the
coverage of the language semantics by the pool.
From the selected target program, \mytextsf{Program Mutator} constructs a new mutated program
and adds it to the program pool.  When specific criteria, such as an iteration limit, are satisfied,
\mytextsf{Target Selector} stops selecting target programs and returns
the program pool as its result.

\subsubsection{Program Mutator}
The main goal of \mytextsf{Program Mutator} is to generate a new program by
mutating a given target program in order to increase the coverage of
the language semantics by the program pool.
If it fails to generate a new program to increase the semantics coverage,
\mytextsf{Target Selector} retries to select a new target program and repeats this
process less than a pre-defined iteration limit.

\subsubsection{Assertion Injector}
Finally, the conformance test generation phase modifies the programs in the pool to generate
conformance tests by injecting appropriate assertions reflecting the
semantics described in the specification.  More specifically,
\mytextsf{Assertion Injector} executes each program in the pool on the mechanized
specification and obtains the final state of its execution.  It then
automatically injects assertions to the program using the final state.

\subsubsection{Bug Localizer}
Then, the second phase executes the conformance tests on $N$ engines and
collects their results.  For each test, if a small number of engines fail,
it reports potential bugs in the engines that fail the test.
Otherwise, it reports potential bugs in the specification.
In addition, its \mytextsf{Bug Localizer} module uses \textit{Spectrum
Based Fault Localization} (SBFL)~\cite{sbfl-survey}, a localization
technique utilizing the coverage and pass/fail results of test cases, to
localize potential bugs.

\section{$N$+1-version Differential Testing for JavaScript}
\label{sec:application}
We actualize $N$+1-version differential testing for the JavaScript programming
language as $\tool$, which uses modern JavaScript engines and ECMAScript.

\subsection{Seed Synthesizer}

$\tool$ synthesizes seed programs using two synthesizers.

\subsubsection{Non-Recursive Synthesizer}
The first synthesizer aims to cover as many syntax cases as possible
in two steps: 1) to find the shortest string for each non-terminal
and 2) to synthesize JavaScript programs using the shortest strings.
For presentation brevity, we explain simple cases like terminals and non-terminals,
but the implementation supports the extended grammar of
ECMAScript such as parametric non-terminals, conditional alternatives,
and special terminal symbols.

\begin{algorithm}[t]
  \caption{Worklist-based Shortest String}
  \label{alg:short-string}
  \DontPrintSemicolon
  \SetKwProg{Fn}{Function}{:}{}
  \SetKwFunction{shortestStrings}{shortestStrings}
  \SetKwFunction{update}{update}
  \SetKwFunction{propagate}{propagate}
  \KwIn{$\ruleset$ - syntax reduction rules}
  \KwOut{$M$ - map from non-terminals to shortest strings derivable
    from them}
  \Fn{\shortestStrings{$\ruleset$}} {
    $M = \varnothing, \worklist = \text{a queue that contains}\ \ruleset$\;
    \While{$\worklist \neq \varnothing$} {
      $\text{pop}\ (A, \alpha) \gets \worklist$\;
      \lIf{$\update(A, \alpha)$} {
        $\propagate(\worklist, \ruleset, A)$}
    }
  }
  \Fn{\update{$A, \alpha$}} {
    $str = \text{an empty string}$\;
    \ForAll{$s \in \alpha$} {
      \lIf{$s\ \text{is a terminal}\ t$} {
        $str = str + t$
      }
      \ElseIf{$s\ \text{is a non-terminal}\ A' \wedge A' \in M$} {
        $str = str + M[A']$
      }
      \lElse {
        \Return false
      }
    }
    \lIf{$\exists M[A] \wedge \norm{str} \geq \norm{M[A]}$} {
      \Return false
    }
    $M[A] = str$\;
    \Return true\;
  }
  \Fn{\propagate{$\worklist, \ruleset, A$}} {
    \ForAll{$(A', \alpha') \in \ruleset$} {
      \lIf{$A \in \alpha'$} {
        $\text{push}\ (A', \alpha') \rightarrow \worklist$
      }
    }
  }
\end{algorithm}

The \mytextsf{shortestStrings} function in Algorithm~\ref{alg:short-string} shows the first step.
We modified McKenize's algorithm~\cite{cfg-gen} that finds random
strings to find the shorted string.  It takes syntax reduction rules
$\ruleset$, a set of pairs of non-terminals and alternatives, and
returns a map $M$ from non-terminals to shortest strings derivable from them.
It utilizes a worklist $W$, a queue structure that includes syntax reduction rules
affected by updated non-terminals.
The function initializes the worklist $W$ with all the syntax reduction rules $\ruleset$.
Then, for a syntax reduction rule $(A, \alpha)$, it updates the
map $M$ via the \mytextsf{update} function, and propagates updated
information via the \mytextsf{propagate} function.
The \mytextsf{update} function checks whether a given alternative $\alpha$ of
a non-terminal $A$ can derive a string shorter than the current
shortest one using the current map $M$.
If possible, it stores the mapping from the non-terminal $A$ to the
newly found shortest string in $M$ and invokes \mytextsf{propagate}.
The \mytextsf{propagate} function finds all the syntax reduction rules
whose alternatives contain the updated non-terminal $A$ 
and inserts them into $W$.  The \mytextsf{shortestStrings} function
repeats this process until the worklist $W$ becomes empty.

\begin{algorithm}[t]
  \caption{Non-Recursive Synthesize}
  \label{alg:non-rec-synthesize}
  \DontPrintSemicolon
  \SetKwProg{Fn}{Function}{:}{}
  \SetKwFunction{nonRecSynthesize}{nonRecSynthesize}
  \SetKwFunction{getProd}{getProd}
  \SetKwFunction{getAlt}{getAlt}
  \KwIn{$\ruleset$ - syntax reduction rules, $S$ - start symbol}
  \KwOut{$D$ - set of strings derivable from $S$}
  \SetKwBlock{Begin}{function}{end function}
  \Fn{\nonRecSynthesize{$\ruleset, S$}} {
    $V = \varnothing, M = \shortestStrings(\ruleset)$\;
    \Return $\getProd(M, V, \ruleset, S)$\;
  }
  \Fn{\getProd{$M, V, \ruleset, A$}} {
    \lIf{$A \in V$} {
      \Return $\{ M[A] \}$
    }
    $D = \varnothing, V = V \cup \{ A \}$\;
    \ForAll{$(A', \alpha) \in \ruleset\ \text{s.t.}\ A' = A$} {
      $D = D \cup \getAlt(M, V, \ruleset, A, \alpha)$\;
    }
    \Return $D$\;
  }
  \Fn{\getAlt{$M, V, \ruleset, A, \alpha$}} {
    $L = \text{an empty list}$\;
    \ForAll{$s \in \alpha$} {
      \If{$s\ \text{is a terminal}\ t$} {
        $\text{append}\ (\{ t \}, t)\ \text{to}\ L$\;
      }
      \ElseIf{$s\ \text{is a non-terminal}\ A'$} {
        $\text{append}\ (\getProd(M, V, \ruleset, A'), M[A])\ \text{to}\ L$}
    }
    $D =$ point-wise concatenation of first elements of pairs in $L$ using
    second elements as default ones.\;
    \Return $D$\;
  }
\end{algorithm}

Using shortest strings derivable from non-terminals, 
the \mytextsf{nonRecSynthesize} function in Algorithm~\ref{alg:non-rec-synthesize}
synthesize programs.  It takes syntax reduction rules $\ruleset$ and a start symbol $S$.
For the first visit with a non-terminal $A$, the \mytextsf{getProd} function returns
strings generated by \mytextsf{getAlt} with alternatives of the non-terminal $A$.
For an already visited non-terminal $A$, it returns the single shortest string $M[A]$.
The \mytextsf{getAlt} function takes a non-terminal $A$ with an alternative
$\alpha$ and returns a set of strings derivable from $\alpha$ via
point-wise concatenation of strings derived by symbols of $\alpha$.
When the numbers of strings derived by symbols are different,
it uses the shortest strings derived by symbols as default strings.

\begin{figure}[t]
  \centering
  \includegraphics[width=0.31\textwidth]{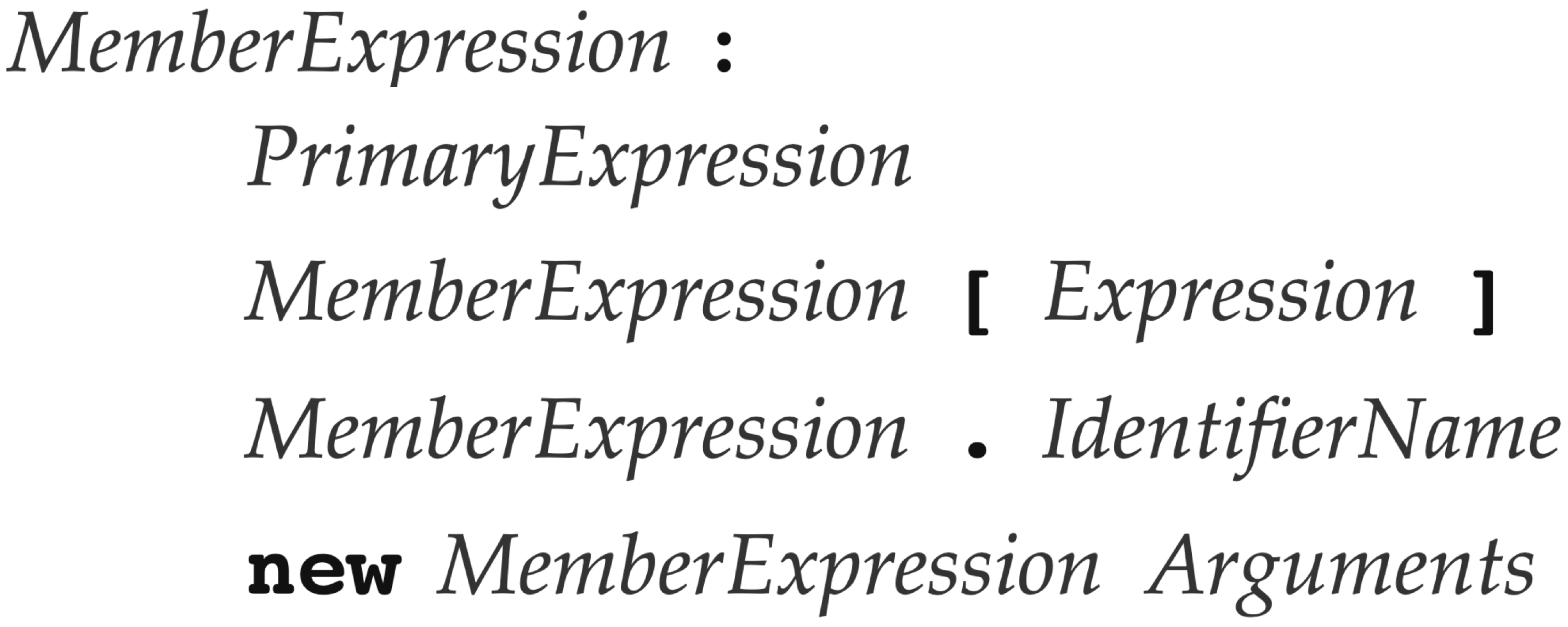}
  \caption{The \textit{MemberExpression} production in ES11}
  \label{fig:prod-example}
  \vspace*{-1em}
\end{figure}

For example, Figure~\ref{fig:prod-example} shows a simplified
\textit{MemberExpression} production in ES11.  For the first step, we find
the shortest string for each non-terminal: \code{()} for \textit{Arguments} and
\code{x} for the other non-terminals.  Note that we use pre-defined shortest
strings for identifiers and literals such as \code{x} for identifiers and
\code{0} for numerical literals.  In the next step, we synthesize strings
derivable from \textit{MemberExpression}.  The first alternative is a
single non-terminal \textit{PrimaryExpression}, which is never visited.
Thus, it generates all cases of \textit{PrimaryExpression}.  The fourth
alternative consists of one terminal \code{new} and two non-terminals
\textit{MemberExpression} and \textit{Arguments}.  Because \textit{MemberExpression}
is already visited, it generates a single shortest string \code{x}.
For the first visit of \textit{Arguments}, it generates all cases:
\code{()}, \code{(x)}, \code{(...x)}, and \code{(x,)}.
Note that the numbers of strings generated for symbols are different.
In such cases, we use the shortest strings for symbols like
\code{x} for \textit{MemberExpression} as follows:

\vspace*{-.5em}
\begin{figure}[H]
  \centering
  \includegraphics[width=0.46\textwidth]{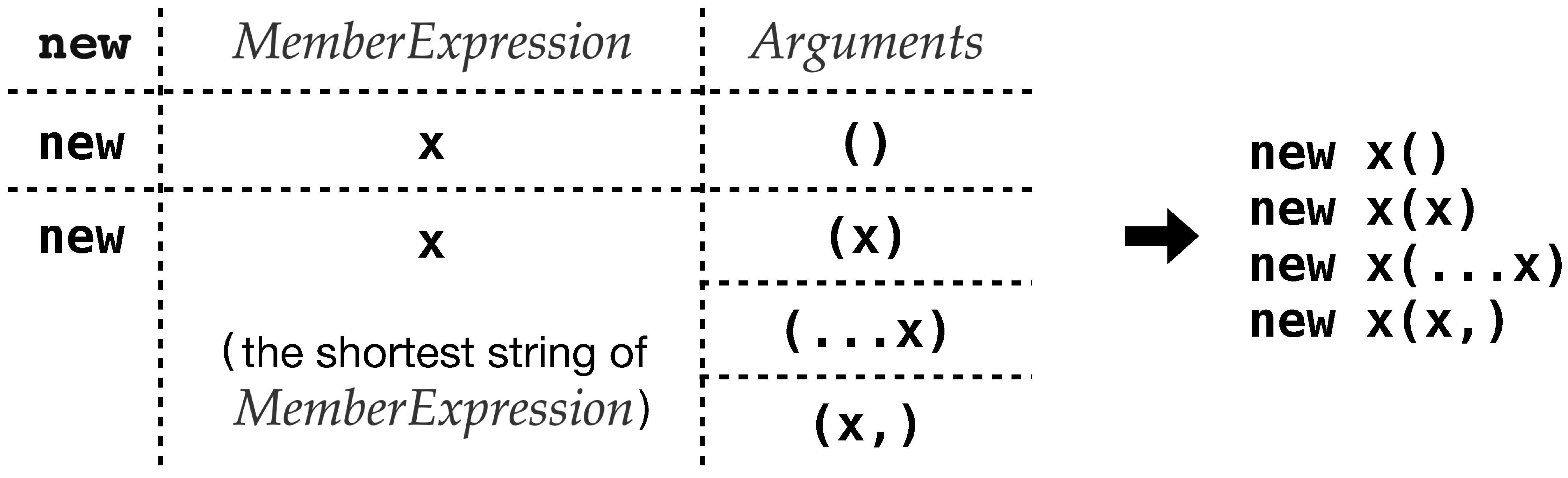}
\end{figure}
\vspace*{-.5em}

\subsubsection{Built-in Function Synthesizer}

JavaScript supports diverse built-in functions for primitive values and built-in objects.
To synthesize JavaScript programs that invoke built-in functions,
we extract the information of each built-in function from the mechanized ECMAScript.
We utilize the \code{Function.prototype.call} function to invoke
built-in functions to easily handle the \code{this} object in \mytextsf{Program Mutator};
we use a corresponding object or \code{null} as the \code{this} object by default.
In addition, we synthesize function calls with optional and variable number of arguments
and built-in constructor calls with the \code{new} keyword.

Consider the following \code{Array.prototype.indexOf} function for
JavaScript array objects that have a parameter \textit{searchElement}
and an optional parameter \textit{fromIndex}:

\vspace*{-.5em}
\begin{figure}[H]
  \centering
  \includegraphics[width=0.4\textwidth]{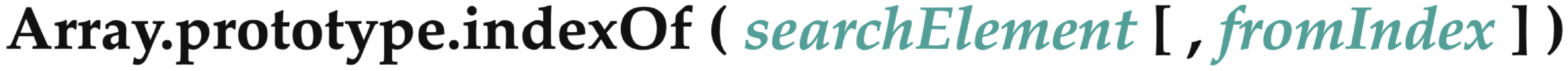}
\end{figure}
\vspace*{-.5em}

\noindent
the synthesizer generates the following calls with an array object or \code{null} as the \code{this}
object as follows:
\begin{lstlisting}[style=myJSstyle]
  Array.prototype.indexOf.call(new Array(), 0);
  Array.prototype.indexOf.call(new Array(), 0, 0);
  Array.prototype.indexOf.call(null, 0);
  Array.prototype.indexOf.call(null, 0, 0);
\end{lstlisting}
Moreover, \code{Array} is a built-in function and
a built-in constructor with a variable number of arguments.
Thus, we synthesize the following six programs for \code{Array}:
\begin{lstlisting}[style=myJSstyle]
    Array();      Array(0);      Array(0, 0);
    new Array();  new Array(0);  new Array(0, 0);
\end{lstlisting}

\subsection{Target Selector}
From the synthesized programs, \mytextsf{Target Selector} selects a
target program to mutate to increase the semantics coverage of the
program pool.  Consider the \textbf{Abstract Equality Comparison} algorithm 
in Figure~\ref{fig:example}(a) again where the first step has the condition
``If Type($x$) is the same as Type($y$).'' Assuming that the current pool
has the following three programs:
\begin{lstlisting}[style=myJSstyle]
           1 + 2;  true == false;  0 == 1;
\end{lstlisting}
because later two programs that perform comparison have values of the same type,
the pool covers only the true branch of the condition in the algorithm.
To cover its false branch, \mytextsf{Target Selector} selects any program
that covers the true branch like \code{true == false;} and \mytextsf{Program Mutator}
mutates it to \code{42 == false;} for example.  Then, since the mutated
program covers the false branch, the pool is extended as follows:
\begin{lstlisting}[style=myJSstyle]
   1 + 2;  true == false;  0 == 1;  42 == false;
\end{lstlisting}
which now covers more steps in the algorithm.
This process repeats until the semantics coverage converges.

\subsection{Program Mutator}
$\tool$ increases the semantics coverage of the program pool by mutating programs
using five mutation methods randomly.

\subsubsection{Random Mutation}
The first na\"ive method is to randomly select a statement, a
declaration, or an expression in a given program and to replace it
with a randomly selected one from a set of syntax trees generated by
the non-recursive synthesizer.
For example, it may mutate a program \code{var x = 1 + 2;} by replacing
its random expression \code{1} with a random expression \code{true} producing
\code{var x = true + 2;}.

\subsubsection{Nearest Syntax Tree Mutation}
The second method targets uncovered branches in abstract algorithms.
When only one branch is covered by a program, it finds the nearest
syntax tree in the program that reaches the branch in the algorithm, and
replaces the nearest syntax tree with a random syntax tree derivable
from the same syntax production.  For example, consider the following JavaScript program:
\begin{lstlisting}[style=myJSstyle]
                var x = "" + (1 == 2);
\end{lstlisting}
While it covers the false branch of the first step of
\textbf{Abstract Equality Comparison} in Figure~\ref{fig:example}(a),
assume that no program in the program pool can cover its true branch.
Then, the mutator targets this branch, finds its nearest syntax tree
\code{1 == 2} in the program, and replaces it with a random syntax tree.

\subsubsection{String Substitutions}
We collect all string literals used in conditions of the algorithms in ES11
and use them for random expression substitutions.
Because most string literals in the specification represent corner cases
such as \code{-0}, \code{Infinity}, and \code{NaN}, they are necessary for mutation
to increase the semantics coverage.  For example, the semantics of the
[[DefineOwnProperty]] internal method of array exotic objects depends
on whether the value of its parameter \code{P} is \code{"length"} or not.

\subsubsection{Object Substitutions}
We also collect string literals and symbols used as arguments of object
property access algorithms in ES11, randomly generate objects using them,
and replace random expressions with the generated objects.
Because some abstract algorithms in the specification access object properties
using \textbf{HasProperty}, \textbf{GetMethod}, \textbf{Get}, and
\textbf{OrdinaryGetOwnProperty}, 
objects with such properties are necessary for mutation to achieve high coverage.  
Thus, the mutator mutates a randomly selected expression in a program with a
randomly generated object that has properties whose keys are from
collected string literals and symbols.

\subsubsection{Statement Insertion}
To synthesize more complex programs, the mutator inserts random
statements at the end of randomly selected blocks like
top-level code and function bodies.  We generate random
statements using the non-recursive synthesizer with pre-defined
special statements.  The special statements are control diverters, which
have high chances of changing execution paths, such as function calls,
\code{return}, \code{break}, and \code{throw} statements.  The mutator
selects special statements with a higher probability than the statements
randomly synthesized by the non-recursive synthesizer.

\subsection{Assertion Injector}

After generating JavaScript programs, \mytextsf{Assertion Injector} injects
assertions to them using their final states as specified in ECMAScript.
It first obtains the final state of a given program from the
mechanized specification and injects seven kinds of assertions in the beginning of the program.
To check the final state after executing all asynchronous jobs,
we enclose assertions with \code{setTimeout} to wait 100 ms when a
program uses asynchronous features such as \code{Promise} and \code{async}:
\begin{lstlisting}[style=myJSstyle]
  ... /* a given program */
  setTimeout(() => { ... /* assertions */ }, 100)
\end{lstlisting}

\subsubsection{Exceptions}

JavaScript supports both internal exceptions like \code{SyntaxError} and
\code{TypeError} and custom exceptions with the keyword \code{throw}.
Note that catching such exceptions using the \code{try-catch} statement
may change the program semantics.  For example, the following does not throw any exception:
\begin{lstlisting}[style=myJSstyle]
    var x; function x() {}
\end{lstlisting}
but the following:
\begin{lstlisting}[style=myJSstyle]
    try { var x; function x() {} } catch (e) {}
\end{lstlisting}
throws \code{SyntaxError} because declarations of a variable and a function
with the same name are not allowed in \code{try-catch}.

To resolve this problem, we exploit a comment in the first line of a program.
If the program throws an internal exception, we tag its name in the comment.
Otherwise, we tag \code{// Throw} for a custom exception and \code{// Normal} for
normal termination.  Using the tag in the comment, $\tool$ checks the
execution result of a program in each engine.

\subsubsection{Aborts}

The mechanized semantics of ECMAScript can abort due to unspecified cases.
For example, consider the following JavaScript program:
\begin{lstlisting}[style=myJSstyle]
    var x = 42; x++;
\end{lstlisting}
The postfix increment operator (\code{++}) increases the number value
stored in the variable \code{x}.  However, because of a typo in the
\textbf{Evaluation} algorithm for such update expressions in ES11, the
behavior of the program is not defined in ES11.  To represent this situation in
the conformance test, we tag \code{Abort} in the comment as follows:
\begin{lstlisting}[style=myJSstyle]
    // Abort
    var x = 42; x++;
\end{lstlisting}

\subsubsection{Variable Values}

We inject assertions that compare the values of variables with expected values.
To focus on variables introduced by tests, we do not
check the values of pre-defined variables like built-in objects.
For numbers, we distinguish \code{-0} from \code{+0} using division by zero because \code{1/-0} and
\code{1/+0} produce negative and positive infinity values, respectively.
The following example checks whether the value of \code{x} is \code{3}:
\begin{lstlisting}[style=myJSstyle]
    var x = 1 + 2;
    $assert.sameValue(x, 3);
\end{lstlisting}

\subsubsection{Object Values}

To check the equality of object values, we keep a representative path for each object.
If the injector meets an object for the first time, it keeps the current path of the object 
as its representative path and injects assertions for the properties of the object.
Otherwise, the injector adds assertions to compare the values of the
objects with the current path and the representative path.  In
the following example:
\begin{lstlisting}[style=myJSstyle]
    var x = {}, y = {}, z = { p: x, q: y };
    $assert.sameValue(z.p, x);
    $assert.sameValue(z.q, y);
\end{lstlisting}
because the injector meets two different new objects
stored in \code{x} and \code{y}, it keeps the paths \code{x} and \code{y}.
Then, the object stored in \code{z} is also a new object but its
properties \code{z.p} and \code{z.q} store already visited objects values.
Thus, the injector inserts two assertions that check whether \code{z.p} and \code{x} have
the same object value and \code{z.q} and \code{y} as well.
To handle built-in objects, we store all the paths of built-in objects in advance.

\subsubsection{Object Properties}

Checking object properties involves checking four attributes for each property.
We implement a helper \code{\$verifyProperty} to check the attributes of each property for each object.
For example, the following code checks the attributes of the property of \code{x.p}:
\begin{lstlisting}[style=myJSstyle]
    var x = { p: 42 };
    $verifyProperty(x, "p", {
        value: 42.0,      writable: true,
        enumerable: true, configurable: true
    });
\end{lstlisting}

\subsubsection{Property Keys}

Since ECMAScript 2015 (ES6), the specification defines orders between property keys in objects.
We check the order of property keys by \code{Reflect.ownKeys},
which takes an object and returns an array of the object's property keys.
We implement a helper \code{\$assert.compareArray} that takes two
arrays and compares their lengths and contents.
For example, the following program checks the property keys and their order of the object in \code{x}:
\begin{lstlisting}[style=myJSstyle]
var x = {[Symbol.match]: 0, p: 0, 3: 0, q: 0, 1: 0}
$assert.compareArray(
    Reflect.ownKeys(x),
    ["1", "3", "p", "q", Symbol.match]
);
\end{lstlisting}

\subsubsection{Internal Methods and Slots}

While internal methods and slots of JavaScript objects are generally inaccessible by users,
the names in the following are accessible by indirect getters:
\[
\small
  \begin{array}{l|l}
    \telem{c|}{Name}   & \telem{c}{Indirect Getter}\\\hline
    \text{[[Prototype]]}  & \code{Object.getPrototypeOf(x)}\\\hline
    \text{[[Extensible]]} & \code{Object.isExtensible(x)}\\\hline
    \text{[[Call]]}       & \code{typeof f === "function"}\\\hline
    \text{[[Construct]]}  & \code{Reflect.construct(function()\{\},[],x)}
  \end{array}
\]

The internal slot [[Prototype]] represents the prototype object of an object,
which is available by a built-in function \code{Object.getPrototypeOf}.
The internal slot [[Extensible]] is also available by a built-in function \code{Object.isExtensible}.
The internal methods [[Call]] and [[Construct]] represent whether a given object is
a function and a constructor, respectively.  Because the methods are not JavaScript values,
we simply check their existence using helpers \code{\$assert.callable}
and \code{\$assert.constructable}.  For [[Call]], we use the \code{typeof} operator because it returns
\code{"function"} if and only if a given value is an object with the [[Call]] method.
For [[Construct]] method, we use the \code{Reflect.construct} 
built-in function that checks the existence of the [[Construct]] methods and invokes it.
To avoid invoking [[Construct]] unintentionally, we call \code{Reflect.construct} with
a dummy function \code{function()\{\}} as its first argument and 
a given object as its third argument.  For example, the following code shows
how the injector injects assertions for internal methods and slots:
\begin{lstlisting}[style=myJSstyle]
  function f() {}
  $assert.sameValue(Object.getPrototypeOf(f),
                    Function.prototype);
  $assert.sameValue(Object.isExtensible(x), true);
  $assert.callable(f);
  $assert.constructable(f);
\end{lstlisting}

\subsection{Bug Localizer}

The bug detection and localization phase uses the execution results of
given conformance tests on multiple JavaScript engines.
If a small number of engines fail in running a specific conformance test,
the engines may have bugs causing the test failure.
If most engines fail for a test, the test may be incorrect,
which implies a bug in the specification.

When we have a set of failed test cases that may contain bugs of an engine or a
specification, we classify the test cases using their failure
messages and give ranks between possible buggy program elements to localize the bug.
We use Spectrum Based Fault Localization (SBFL)~\cite{sbfl-survey},
which is a ranking technique based on likelihood of being faulty for each
program element.  We use the following formula called $ER1_b$,
which is one of the best SBFL formulae theoretically analyzed by Xie et al.~\cite{er1b}:
\[
  {n_{\mbox{\emph{\scriptsize ef}}}} -
  {
    {n_{\mbox{\emph{\scriptsize ep}}}}
    \over
    {n_{\mbox{\emph{\scriptsize ep}}} + n_{\mbox{\emph{\scriptsize np}}} + 1}
  }
\]
where $n_{ef}$, $n_{ep}$ , $n_{nf}$, and $n_{np}$ represent the number of test
cases; subscripts ${}_e$ and ${}_n$ respectively denote whether a test case touches a
relevant program element or not, and subscripts ${}_f$ and ${}_p$
respectively denote whether the test case is failed or passed.

We use abstract algorithms of ECMAScript as program elements used for SBFL.
To improve the localization accuracy, we use method-level aggregation~\cite{fluccs}.
It first calculates SBFL scores for algorithm steps and aggregates
them up to algorithm-level using the highest score among those from steps of each algorithm.

\section{Evaluation}\label{sec:eval}

\begin{figure*}[t]
  \centering
  \begin{subfigure}[t]{0.48\textwidth}
    \includegraphics[width=\textwidth]{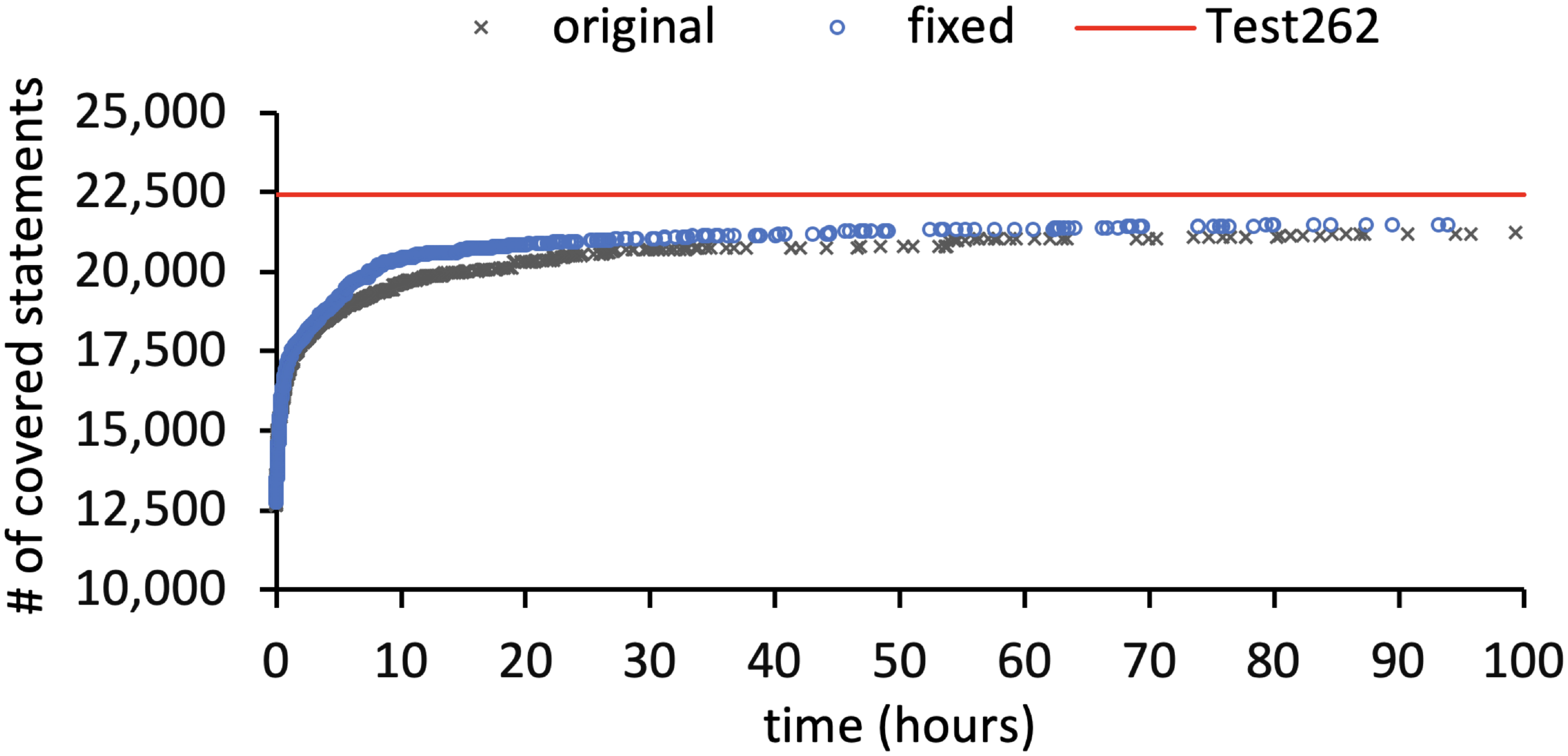}
    \caption{Statement coverage}
    \label{fig:stmt-coverage}
  \end{subfigure}
  \quad
  \begin{subfigure}[t]{0.48\textwidth}
    \includegraphics[width=\textwidth]{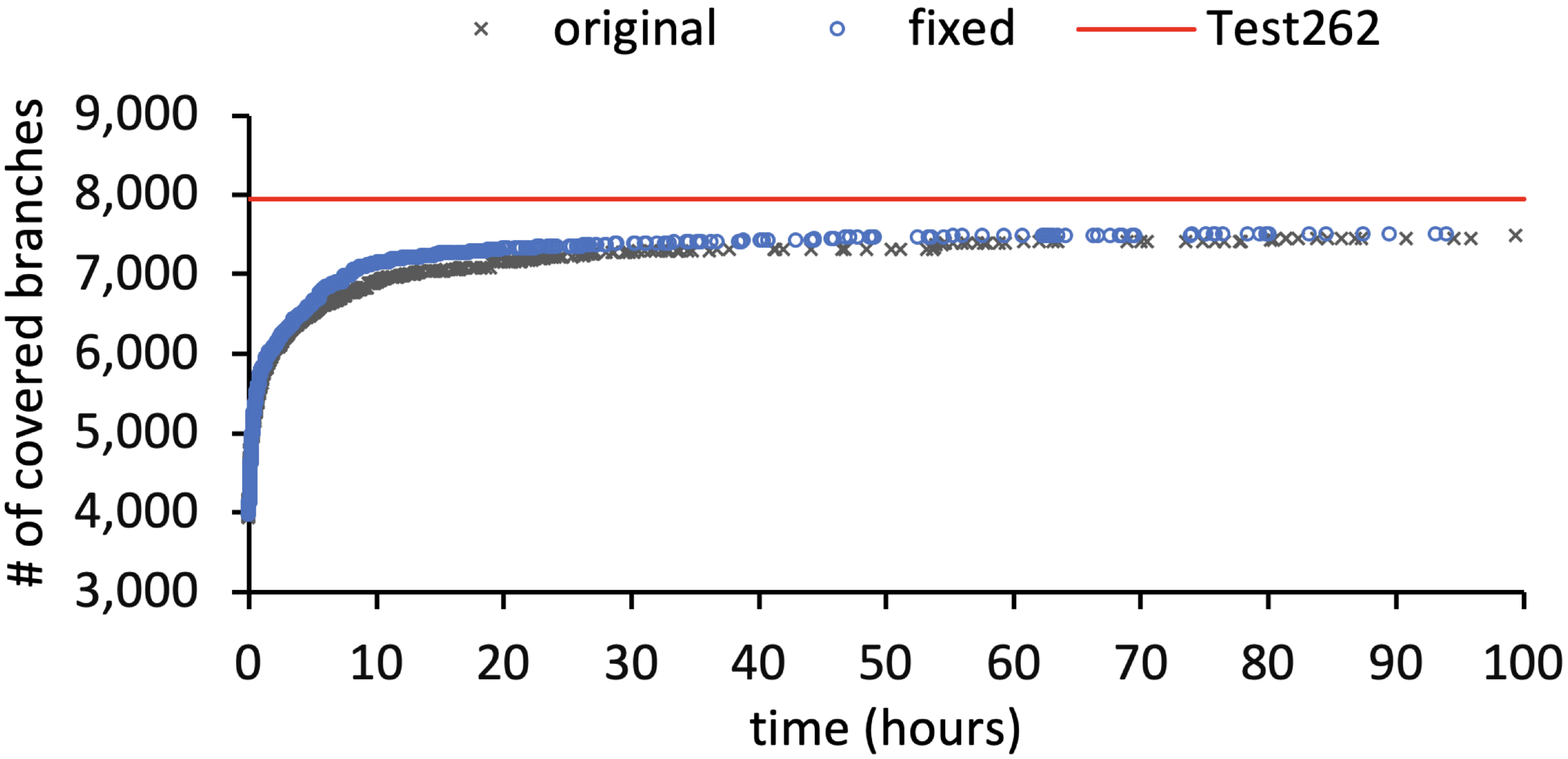}
    \caption{Branch coverage}
    \label{fig:branch-coverage}
  \end{subfigure}
  \caption{The semantics coverage changes during the test generation phase}
  \label{fig:sem-coverage}
  \vspace*{-1em}
\end{figure*}

To evaluate $\tool$ that performs $N$+1-version differential testing of JavaScript
engines and its specification, we applied the tool to four JavaScript engines that
fully support modern JavaScript features and the latest specification,
ECMAScript 2020 (ES11, 2020).  Our experiments use the following four JavaScript
engines, all of which support ES11:
\begin{itemize}
  \item \textbf{V8(v8.3)\footnote{https://v8.dev/}:} An open-source high-performance engine
    for JavaScript and WebAssembly developed by Google~\cite{v8}
  \item \textbf{GraalJS(v20.1.0)\footnote{https://github.com/graalvm/graaljs\#current-status}:} A JavaScript implementation built on
    GraalVM~\cite{graaljs}, which is a Java Virtual Machine (JVM) based on
    HotSpot/OpenJDK developed by Oracle
  \item \textbf{QuickJS(2020-04-12)\footnote{https://bellard.org/quickjs/}:} A small and embedded JavaScript engine developed by
    Fabrice Bellard and Charlie Gordon~\cite{qjs}
  \item \textbf{Moddable XS(v10.3.0)\footnote{https://blog.moddable.com/blog/xs10/}:} A JavaScript engine at the center of the Moddable
    SDK~\cite{xs}, which is a combination of development tools and runtime
    software to create applications for micro-controllers
\end{itemize}
To extract a mechanized specification from ECMAScript, we utilize the tool
$\jiset$, which is a JavaScript IR-based semantics extraction
toolchain, to automatically generate a JavaScript interpreter from ECMAScript.
To focus on the core semantics of JavaScript, we consider only the semantics of strict mode
JavaScript code that pass syntax checking including the EarlyError rules.  To
filter out JavaScript code that are not strict or fail syntax checking,
we utilize the syntax checker of the most reliable JavaScript engine, V8.
We performed our experiments on a machine equipped with 4.0GHz Intel(R) Core(TM)
i7-6700k and 32GB of RAM (Samsung DDR4 2133MHz 8GB*4).  We evaluated $\tool$
with the following four research questions:
\begin{itemize}
\item {\bf RQ1 (Coverage of Generated Tests)} Is the semantics
coverage of the tests generated by $\tool$ comparable to that of Test262,
the official conformance test suite for ECMAScript, which is manually written?
\item {\bf RQ2 (Accuracy of Bug Localization)} Does $\tool$ localize bug locations
accurately?
\item {\bf RQ3 (Bug Detection in JavaScript Engines)} How many
bugs of four JavaScript engines does $\tool$ detect?
\item {\bf RQ4 (Bug Detection in ECMAScript)} How many
bugs of ES11 does $\tool$ detect?
\end{itemize}

\subsection{Coverage of Generated Tests}

$\tool$ generates the seed programs via \mytextsf{Seed Synthesizer},
which synthesizes 1,125 JavaScript programs in about 10
seconds and covers 97.78\% (397/406) of reachable 
alternatives in the syntax productions of ES11.
Among them, we filtered out 602 programs that do not increase
the semantics coverage and started the mutation iteration with 519 programs.
Figure~\ref{fig:sem-coverage} shows the change of
semantics coverage of the program pool during the iterative process in 100 hours.
The left and right graphs present the statement and branch coverages,
respectively, and the top red line denotes the coverage of Test262.
We generated conformance tests two times before and after fixing bugs detected
by $\tool$ because the specification bugs affected the semantics coverage.
In each graph, dark gray X marks and blue O marks denote the semantics coverage
of generated tests before and after fixing bugs.
The semantics that we target in ES11 consists of 1,550 algorithms with 24,495
statements and 9,596 branches.
For the statement coverage, Test262 covers 22,440 (91.61\%) statements.
The initial program pool covers 12,768 (52.12\%) statements
and the final program pool covers 21,230 (86.67\%) and
21,482 (87.70\%) statements before and after fixing bugs, respectively.
For the branch coverage, Test262 covers 7,956 (82.91\%) branches.
The initial program pool covers 3,987 (41.55\%) branches
and the final program pool covers 7,480 (77.95\%) and
7,514 (78.30\%) branches before and after fixing bugs, respectively.

\begin{table}
  \caption{Number of generated programs and covered branches of mutation methods}
  \label{table:mutation-method}
  \vspace*{-1em}
  \small
  \[
    \begin{array}{l?r|r}
      \telembf{c?}{Mutation Method}      & \telembf{c|}{Program}  & \telembf{c}{Branch (Avg.)}\\\toprule\\[-1.4em]
      \text{Nearest Syntax Tree Mutation} & 459                   & 1,230 (2.68)\\\hline
      \text{Random Mutation}              & 337                   & 1,153 (3.42)\\\hline
      \text{Statement Insertion}          & 209                   & 650   (3.11)\\\hline
      \text{Object Substitution}          & 169                   & 491   (2.91)\\\hline
      \text{String Substitution}          & 3                     & 3     (1.00)\\\hline
      \hline
      \telembf{c?}{Total}                 & 1,177                 & 3,527 (3.00)\\
    \end{array}
  \]
  \vspace*{-3em}
\end{table}

Table~\ref{table:mutation-method} shows the number of synthesized programs and covered
branches for each mutation method during the test generation phase.  In total,
$\tool$ successfully synthesize 1,177 new programs that cover 3,527
more branches than the initial program pool.  Among five mutation methods, the
nearest syntax tree mutation is the most contributed method (459
programs and 1,230 covered branches) and the least one is the string
substitution (3 programs and 3 covered branches).  On average,
3.00 branches are covered by a new program.

Finally, $\tool$ generates 1,700 JavaScript programs and their average number of
lines is 2.01.  After injecting assertions, their average number of lines
becomes 8.45.  Compared to Test262, the number of generated tests are much
smaller and their number of lines are also shorter than those of tests in
Test262.  Test262 provides 16,251 tests for the same range of semantics and
their average number of lines is 49.67.

\subsection{Accuracy of Bug Localization}

\begin{figure}[t]
  \centering
  \includegraphics[width=0.48\textwidth]{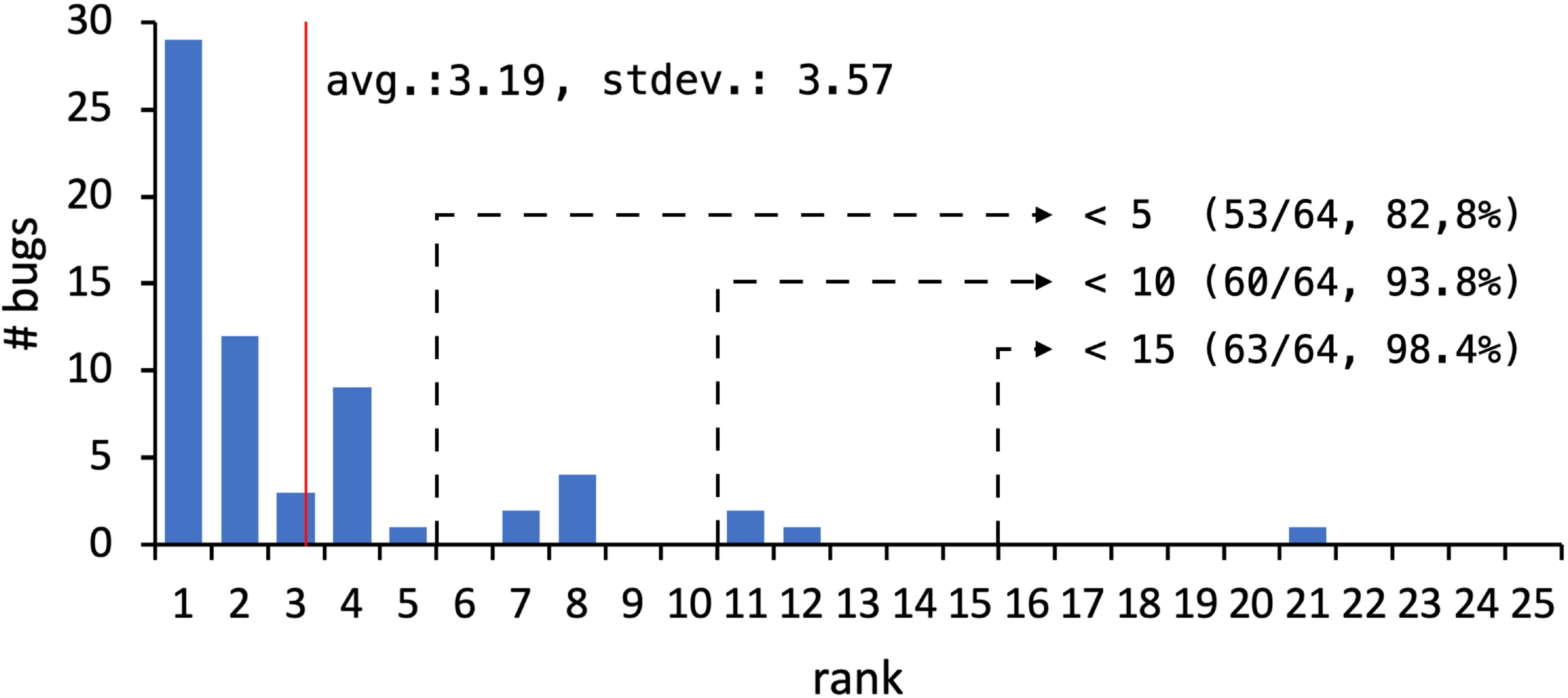}
  \caption{Ranks of algorithms that caused the bugs detected by $\tool$}
  \label{fig:localize}
  \vspace*{-1em}
\end{figure}

\setcounter{table}{2}
\begin{table*}[t]
  \centering
  \caption{Specification bugs in ECMAScript 2020 (ES11) detected by $\tool$}
  \label{table:spec-bug}
  \vspace*{-.5em}
  \small
  \begin{tabular}{@{}c@{~}?c|@{~}c@{~}|l|c|c|@{~}c@{~}|@{~}c@{~}|@{~}r@{}}
    \telembf{@{}c?}{\bf Name} &
    \telembf{c}{\bf Feature} &
    \telembf{@{}c@{~}}{\bf \#} &
    \telembf{c}{\bf Description} &
    \telembf{@{~}c@{~}}{\bf Assertion} &
    \telembf{@{~}c}{\bf Known} &
    \telembf{@{}c}{\bf Created} &
    \telembf{@{}c}{\bf Resolved} &
    \telembf{@{}c@{~}}{\bf Existed} \\\toprule\\[-1.4em]

    ES11-1 &
    \text{Function} &
    12 &
    \makecell[l]{Wrong order between property keys for functions} &
    \mytextsf{Key} &
    O &
    2019-02-07 &
    2020-04-11 &
    429 days \\\hline

    ES11-2 &
    \text{Function} &
    8 &
    \makecell[l]{Missing property \code{name} for anonymous functions} &
    \mytextsf{Key} &
    O &
    2015-06-01 &
    2020-04-11 &
    1,776 days \\\hline

    ES11-3 &
    \text{Loop} &
    1 &
    \makecell[l]{Returning iterator objects instead of iterator records\\
      in \textbf{ForIn/OfHeadEvaluation} for \code{for-in} loops} &
    \mytextsf{Exc} &
    O &
    2017-10-17 &
    2020-04-30 &
    926 days \\\hline

    ES11-4 &
    \text{Expression} &
    4 &
    \makecell[l]{Using the wrong variable \code{oldvalue} instead of\\
      \code{oldValue} in \textbf{Evaluation} of \textit{UpdateExpression}} &
    \mytextsf{Abort} &
    O &
    2019-09-27 &
    2020-04-23 &
    209 days \\\hline

    ES11-5 &
    \text{Expression} &
    1 &
    \makecell[l]{Unhandling abrupt completion\\
      in \textbf{Abstract Equality Comparison}} &
    \mytextsf{Exc} &
    O &
    2015-06-01 &
    2020-04-28 &
    1,793 days \\\hline

    ES11-6 &
    \text{Object} &
    1 &
    \makecell[l]{Unhandling abrupt completion in \textbf{Evaluation} of\\
      \textit{PropertyDefinition} for object literals} &
    \mytextsf{Exc} &
    X &
    2019-02-07 &
    TBD &
    TBD
  \end{tabular}
  \vspace*{-0.5em}
\end{table*}

\setcounter{table}{1}
\begin{table}
  \caption{The number of engine bugs detected by $\tool$}
  \label{table:engine-bug}
  \vspace*{-1em}
  \small
  \[
    \begin{array}{l?r|r|r|r|r|r|r?r}
      \telembf{@{}c@{~}?}{Engines} &
      \telemsf{@{~}c@{~}|}{Exc} &
      \telemsf{@{~}c@{~}|}{Abort} &
      \telemsf{@{~}c@{~}|}{Var} &
      \telemsf{@{~}c@{~}|}{Obj} &
      \telemsf{@{~}c@{~}|}{Desc} &
      \telemsf{@{~}c@{~}|}{Key} &
      \telemsf{@{~}c@{~}?}{In} &
      \telembf{@{~}c@{}}{Total}\\\toprule\\[-1.4em]

      \text{V8}           & 0   & 0 & 0 & 0 & 0 & 2   & 0 & 2\\\hline
      \text{GraalJS}      & 6   & 0 & 0 & 0 & 2 & 8   & 0 & 16\\\hline
      \text{QuickJS}      & 3   & 0 & 1 & 0 & 0 & 2   & 0 & 6\\\hline
      \text{Moddable XS}  & 12  & 0 & 0 & 0 & 3 & 5   & 0 & 20\\\hline
      \hline
      \telembf{c?}{Total} & 21  & 0 & 1 & 0 & 5 & 17  & 0 & 44\\
    \end{array}
  \]
  \vspace*{-1.5em}
\end{table}

To detect more bugs using more diverse programs,
we repeated the conformance test generation phase for ten times.
We executed the generated conformance tests on four JavaScript engines
to find bugs in the engines and the specification.
After inferring locations of the bugs in the engines or the specification
based on the majority of the execution results, we manually checked
whether the bugs are indeed in the engines or the specification. 
The following table shows that our method works well:

\begin{table}[H]
  \centering
  \vspace*{-1em}
  \small
  \[
    \begin{array}{l?r|r|r|r?r?r}
      \telembf{c?}{\# Failed Engines} &
      \telembf{c}{1} &
      \telembf{c}{2} &
      \telembf{c}{3} &
      \telembf{c?}{4} &
      \telembf{c?}{Total} &
      \telembf{c}{Average} \\\toprule\\[-1.4em]

      \text{Engine Bugs}        & 38  & 6   & 0   & 0   & 44  & 1.14\\\hline
      \text{Specification Bugs} & 0   & 0   & 10  & 17  & 27  & 3.63\\
    \end{array}
  \]
  \vspace*{-1em}
\end{table}

\noindent
For engine bugs, the average number of engine failures is 1.14
while the average number of failed engines for specification bugs is 3.63.
As we expected, when most engines fail for a test, the specification
may have a bug.

Based on the results of conformance tests on four JavaScript engines, we localized
the specification or engine bugs on the \emph{semantics} of ES11.
Among 71 bugs, we excluded 7 syntax bugs and localized only 64 semantics bugs.
Figure~\ref{fig:localize} shows the ranks of algorithms that caused the semantics bugs.
The average rank is 3.19, and 82.8\% of the algorithms causing the
bugs are ranked less than 5, 93.8\% less than 10, and 98.4\% less than 15.
Note that the location of one bug is ranked 21 because of the limitation of SBFL;
its localization accuracy becomes low for a small number of failed test cases.

\subsection{Bug Detection in JavaScript Engines}
From four JavaScript engines, $\tool$ detected 44 bugs:
2 from V8, 16 from GraalJS,
6 from QuickJS, and 20 from Moddable XS.
Table~\ref{table:engine-bug} presents how many bugs for each assertion are detected
for each engine.  We injected seven kinds of assertions: exceptions
(\mytextsf{Exc}), aborts (\mytextsf{Abort}), variable values (\mytextsf{Var}), object
values (\mytextsf{Obj}), object properties (\mytextsf{Desc}), property keys
(\mytextsf{Key}), and internal methods and slots (\mytextsf{In}).
The effectiveness of bug finding is different for different assertions.
The \mytextsf{Exc} and \mytextsf{Key} assertions detected
engine bugs the most; out of 44 bugs, the former detected 21 bugs
and the latter detected 17 bugs.
\mytextsf{Desc} and \mytextsf{Var} detected 5 and 1 bugs, respectively, but the
other assertions did not detect any engine bugs.

The most reliable JavaScript engine is V8 because $\tool$ found only two bugs and
the bugs are due to specification bugs in ES11.  Because V8 strictly follows the
semantics of functions described in ES11, it also implemented wrong semantics
that led to ES11-1 and ES11-2 listed in Table~\ref{table:spec-bug}.
The V8 team confirmed the bugs and fixed them.

We detected 16 engine bugs in GraalJS and one of them caused an engine
crash.  When we apply the prefix increment operator for \code{undefined}
as \code{++undefined}, GraalJS throws \code{java.lang.IllegalStateException}.
Because it crashes the engine, developers even cannot catch the exception as follows:
\begin{lstlisting}[style=myJSstyle]
    try { ++undefined; } catch(e) { }
\end{lstlisting}
The GraalJS team has been fixing the bugs we reported and
asked whether we plan to publish the conformance test suite,
because the tests generated by $\tool$ detected many semantics bugs that
were not detected by other conformance tests:
\emph{``Right now, we are running Test262 and the V8 and Nashorn
unit test suites in our CI for every change, it might make sense to
add your suite as well.''}

In QuickJS, $\tool$ detected 6 engine bugs, most of which are due to corner cases of
the function semantics.  For example, the following code should throw
a \code{ReferenceError} exception:
\begin{lstlisting}[style=myJSstyle]
    function f (... { x = x }) { return x; } f()
\end{lstlisting}
because the variable \code{x} is not yet initialized when it tries to
read the right-hand side of \code{x = x}.
However, since QuickJS assumes that the initial value of \code{x} is
\code{undefined}, the function call \code{f()} returns \code{undefined}.
The QuickJS team confirmed our bug reports and it has been fixing the bugs.

$\tool$ found the most bugs in Moddable XS; it detected 20 bugs for various
language features such as optional chains, \code{Number.prototype.toString},
iterators of \code{Map} and \code{Set}, and complex assignment patterns.
Among them, optional chains are newly introduced in ES11, which shows that
our approach is applicable to finding bugs in new language features.
We reported all the bugs found, and the Moddable XS team has been
fixing them.  They showed interests in using our test suite:
\emph{``As you know, it is difficult to verify changes because the language specification
is so big. Test262, as great a resource as it is, is not definitive.''}

\subsection{Bug Detection in ECMAScript}
From the latest ECMAScript ES11, $\tool$ detected 27 specification bugs.
Table~\ref{table:spec-bug} summarizes the bugs categorized by their root causes.
Among them, five categories (ES11-1 to ES11-5) were already reported and fixed in the current
draft of the next ECMAScript but ES11-6 was never reported before.
We reported it to TC39; they confirmed it and they will
fix it in the next version, ECMAScript 2021 (ES12).

ES11-1 contains 12 bugs; it is due to a wrong order between property keys of all kinds of
function values such as \code{async} and generator functions, arrow functions, and classes.
For example, if we define a class declaration with a name \code{A}
(\code{class A \{\}}), three properties are defined in the function
stored in the variable \code{A}: \code{length} with a number value \code{0},
\code{prototype} with an object, and \code{name} with a string \code{"A"}.
The problem is the different order of their keys because of
the wrong order of their creation.
From ECMAScript 2015 (ES6), the order between property keys is no
more implementation-dependent but it is related to the creation order of properties.
While the order of property keys in the class \code{A} should be \code{[length, prototype, name]}
according to the semantics of ES11, the order is \code{[length, name, prototype]}
in three engines except V8.  We found that it was already reported as a specification bug;
we reported it to V8 and they fixed it.
This bug was created on February 7, 2019 and TC39 fixed it on April 11, 2020;
the bug lasted for 429 days.

ES11-2 contains 8 bugs that are due to the missing property
\code{name} of anonymous functions.  Until ES5.1, anonymous functions, such as an identity arrow
function \code{x => x}, had their own property \code{name} with an empty string \code{""}.
While ES6 removed the \code{name} property from anonymous functions,
three engines except V8 still create the \code{name} property in anonymous functions.
We also found that it was reported as a specification bug and reported it
to V8, and it will be fixed in V8.

The bug in ES11-3 comes from the misunderstanding of the term ``iterator
object'' and ``iterator record''.  The algorithm \textbf{ForIn/OfHeadEvaluation}
should return an iterator record, which is an implicit record containing only internal slots.
However, In ES11, it returns an iterator object, which is a
JavaScript object with some properties related to iteration.
It causes a \code{TypeError} exception when executing the code \code{for(var x in \{\});} according to
ES11 but all engines execute the code normally without any exceptions.
This bug was resolved by TC39 on April 30, 2020.

ES11-4 contains four bugs caused by a typo for the variable in the
semantics of four different update expressions: \code{x++}, \code{x--},
\code{++x}, and \code{--x}.  In each \textbf{Evaluation} of four kinds of
\textit{UpdateExpression}, there exists a typo \code{oldvalue} in step 3
instead of \code{oldValue} declared in step 2.  $\tool$ could not execute
the code \code{x++} using the semantics of ES11 because of the typo.
For this case, we directly pass the code to \mytextsf{Bug Localizer} to test whether the
code is executable in real-world engines and to localize the bug.
Of course, four JavaScript engines executed the update expressions without any issues
and this bug was resolved by TC39 on April 23, 2020.

Two bugs in ES11-5 and ES11-6 are caused by unhandling of abrupt completions in
abstract equality comparison and property definitions of object literals, respectively.
The bug in ES11-5 was confirmed by TC39 and was fixed on April 28, 2020.
The bug in ES11-6 was a genuine one, and we reported it and received a confirmation
from TC39 on August 18, 2020. The bug will be fixed in the next version, ES12.

\section{Related Work}\label{sec:related}

Our technique is related to three research fields: differential testing,
fuzzing, and fault localization.

\textbf{Differential Testing:} Differential testing~\cite{diff-test} utilizes
multiple implementations as cross-referencing oracles to find semantics bugs.
Researchers applied this technique to various applications domain such as Java
Virtual Machine (JVM) implementations~\cite{diff-jvm}, SSL/TLS certification
validation logic~\cite{nezha,diff-ssl,diff-ssl2}, web
applications~\cite{diff-web}, and binary lifters~\cite{ir-diff-test}.
Moreover, \textsc{Nezha}~\cite{nezha} introduces a guided differential testing tool with
the concept of $\delta$-diversity to efficiently find semantics bugs.
However, they have a fundamental limitation that they cannot test specifications;
they use only cross-referencing oracles and target potential bugs in implementations.
Our $N$+1-version differential testing extends the idea of
differential testing with not only $N$ different 
implementations but also a mechanized specification to test both of them.
In addition, our approach automatically generates conformance tests directly from
the specification.

\textbf{Fuzzing:} Fuzzing is a software testing technique for detecting
security vulnerabilities by generating~\cite{imf,gen-fuzzing,csmith} or
mutating~\cite{mutate-fuzzing,mutate-fuzzing2,mutate-fuzzing3} test inputs.
For JavaScript~\cite{js-hopl} engines,
Patrice et al.~\cite{grammar-whitebox} presented white-box fuzzing using the JavaScript grammar,
Han et al.~\cite{codealchemist} presented CodeAlchemist that generates JavaScript code
snippets based on semantics-aware assembly,
Wang et al.~\cite{superion} presented Superion using Grammar-aware greybox fuzzing,
Park et al.~\cite{die} presented \textsc{Die} using aspect-preserving mutation,
and Lee et al.~\cite{montage} presented Montage using neural network language models (NNLMs).
While they focus on finding security vulnerabilities rather than
semantics bugs, our $N$+1-version differential testing
focuses on finding semantics bugs by comparing multiple
implementations with the mechanized specification,
which was automatically extracted from ECMAScript by $\jiset$.
Note that $\tool$ can also localize not only specification bugs in ECMAScript
but also bugs in JavaScript engines indirectly using the bug locations in ECMAScript.

\textbf{Fault Localization:} To localize detected bugs in ECMAScript, we used
Spectrum Based Fault Localization (SBFL)~\cite{sbfl-survey}, which is a ranking
technique based on likelihood of being faulty for each program element.
Tarantula~\cite{tarantula, tarantula2} was the first tool that supports SBFL
with a simple formula and researchers have developed many formulae~\cite{ample, zoltar,
sbfl-model, effect-sbfl} to increase the accuracy of bug localization.
Sohn and Yoo~\cite{fluccs} introduced a novel approach for fault
localization using code and change metrics via learning of SBFL formulae.
While we utilize a specific formula $ER1_b$ introduced by Xie et al.~\cite{er1b}, we
believe that it is possible to improve the accuracy of bug localization by
using more advanced SBFL techniques.


\section{Conclusion}\label{sec:conclude}
The development of modern programming languages follows the continuous integration (CI) and
continuous deployment (CD) approach to instantly support fast changing user demands.
Such continuous development makes it difficult to find semantics bugs
in both the language specification and its various implementations.
To alleviate this problem, we present $N$+1-version differential testing,
which is the first technique to test both implementations and its specification in tandem.
We actualized our approach for the JavaScript programming language via $\tool$,
using four modern JavaScript engines and the latest version of ECMAScript (ES11, 2020).
It automatically generated 1,700 JavaScript programs with 97.78\% of syntax
coverage and 87.70\% of semantics coverage on ES11.  $\tool$ injected assertions
to the generated JavaScript programs to convert them as conformance tests.
We executed generated conformance tests on four engines that support ES11:
V8, GraalJS, QuickJS, and Moddable XS.  Using the execution results,
we found 44 engine bugs (16 for GraalJS, 6 for QuickJS,
20 for Moddable XS, and 2 for V8) and 27 specification bugs.
All the bugs were confirmed by TC39, the committee of ECMAScript, and
the corresponding engine teams, and they will be fixed in the specification and the engines.
We believe that $\tool$ takes the first step towards co-evolution of
software specifications, tests, and their implementations for CI/CD.

\section*{Acknowledgements}
This work was supported by National Research Foundation of
Korea (NRF) (Grants NRF-2017R1A2B3012020 and 2017M3C4A7068177).

\bibliographystyle{IEEEtran}
\bibliography{ref}

\end{document}